\documentclass[12pt]{article}
\usepackage[pctex32]{graphics}
\begin{document}
\begin{center}
~\\
{\bf  \Large  Spinning String and Giant Graviton in Electric/Magnetic Field Deformed $AdS_3 \times S^3 \times T^4$}
~\\
~\\
                      Wung-Hong Huang\\
                       Department of Physics\\
                       National Cheng Kung University\\
                       Tainan,Taiwan
\\
~
\\
{\bf  ABSTRACT } \end{center}
We  apply the transformation of mixing azimuthal and internal coordinate or mixing time and internal coordinate to the 11D M-theory with a stack of  M2-branes  $\bot$ M2-branes, then, through the mechanism of Kaluza-Klein reduction and a series of the T duality we obtain the corresponding background of a stack of  D1-branes  $\bot$ D5-branes which, in the near-horizon limit,  becomes the magnetic or electric Melvin field deformed $AdS_3 \times S^3 \times T^4$.  We find the giant graviton solution in the deformed spacetime and see that the configuration whose angular momentum is within a finite region could has a fixed size and become more stable than the point-like graviton, in contrast to the undeformed giant graviton which only exists when its angular momentum is a specific value and could have arbitrary size.   We discuss in detail the properties of how the electric/magnetic Melvin field will affect the size of the giant gravitons.  We also adopt an ansatz to find the classical string solutions which are rotating in the deformed $S^3$ with an angular momentum in the rotation plane.  The spinning string and giant graviton solutions we obtained show that the external magnetic/electric flux will increase the solution energy.  Therefore, from the AdS/CFT point of view, the corrections of the anomalous dimensions of operators in the dual field theory will be positive.  Finally, we also see that the spinning string and giant graviton in the near-horizon spacetime of Melvin field deformed D5-branes background have the similar properties as those in the deformed $AdS_3 \times S^3 \times T^4$.
\\
~
\\
~ 
\begin{flushleft}
E-mail:  whhwung@mail.ncku.edu.tw\\
\end{flushleft}

%%%%%%%%%%%%%%%%%%%%%%%

\newpage
\section{Introduction}
The ADS/CFT correspondence [1-2] relates weakly coupled string theories to the strongly coupled field theories.  We hope that ultimately it could be used to understand non-abelian perturbative aspects of QCD.  From the point of view, we must understand the gauge theory/gravity correspondence in the theory with no supersymmetry.   The generalization of AdS/CFT duality to the non-BPS string mode sector has been investigated by semiclassical considerations [2,3], in which it was found that using the novel multi-spin string states one can  carry out  the  precise test of the  AdS/CFT duality in a  non-BPS sector by comparing the  ${\lambda \over J^2 }\ll 1$  expansion of the {\it classical} string energy  with  the corresponding {\it quantum}  anomalous dimensions in  perturbative SYM theory [4-6].  Another interesting configurations are the giant gravitons [7,8] which have also been used in studying of the gauge/gravity correspondence in supersymmetric examples [9-10]. 

  In recent Lunin and Maldacena [11] had demonstrated that certain deformation of the $AdS_{5}\times S^{5}$ background corresponds to a $\beta$-deformation of $N=4$ SYM gauge theory in which the supersymmetry being broken was studied by Leigh and Strassler [12].

As the supersymmetry may be broken under the magnetic or electric flux  it is useful to investigate string theory  in the background with magnetic or electric field deformation.   It is known that the superstring theory in  $AdS_{5}\times S^{5}$ is dual to the $N=4$ SYM theory, we thus had in the previous papers [13] and [14] investigated the semiclassical rotating string and giant graviton solutions in the electric/magnetic field deformed  $AdS_{5}\times S^{5}$ respectively.  In a similar way, it is known that the string theory in $AdS_{3}\times S^{3}$ is  dual to the $1+1$ CFT [1,10], we will therefore in this paper investigate the classical string and giant graviton solutions in the electric/magnetic field deformed  $AdS_{3}\times S^{3}$. 

  Giant graviton first investigated in [7] is a rotating D3-brane in the $AdS_5 \times S^5$ spacetime, which is blowed up to the spherical BPS configuration and has the same energy and quantum number of the point-like graviton.  The configuration is stable only if its angular momentum was less than a critical value of $P_c$.  In [14] we investigated the properties of the giant graviton in the electric/magnetic Melvin geometries of deformed  $AdS_5 \times S^5$ spacetime and found that the deformed giant graviton has lower energy than the point-like graviton.  However, the giant graviton in the electric Melvin field deformed $AdS_5 \times S^5$ spacetime is always unstable and will transit into a point-like graviton, irrespective of its angular momentum.  

  As the giant graviton in the undeformed  $AdS_{3}\times S^{3}\times T^4$  only exists when its angular momentum is a specific value and, moreover, it could have arbitrary size [7,15],  we will in this paper see how the electric/magnetic flux will dramatically change the property.

   In section II we apply the transformation of mixing azimuthal and internal coordinate [16] to the 11D M-theory with a stack of  M2-branes  $\bot$ M2-branes [17] to find  the corresponding spacetime of a stack of  D2-branes  $\bot$ D2-branes with magnetic Melvin flux  in 10 D IIA string theory, after the Kaluza-Klein reduction.  We then perform a series of  T duality [18] to find the corresponding background of a stack of  D1-branes  $\bot$ D5-branes.  In the near-horizon limit the background becomes the magnetic Melvin field deformed $AdS_3 \times S^3 \times T^4$.  Through the same scheme, in section III  we apply the transformation of mixing time and internal coordinate [19] to obtain the electric Melvin field deformed $AdS_3 \times S^3 \times T^4$. 

  In both of section II and III we follow the analyses of [3] to find the spinning string solution and follow the analyses of [15] to consider the giant graviton solution built by a bound state of D1-branes and D5-branes wrapped on the deformed $T^4$ torus.  We obtain the relation between the classical string energy and its angular momentum and find that the external magnetic/electric fluxes will increase the string energy.  Therefore, from the AdS/CFT point of view [3-6], the corrections of the anomalous dimensions of operators in the dual SYM theory will be positive.  The property is the same as that in deformed $AdS_5\times S^{5}$ theory studied in our previous paper [13]. 

 We also in section II and III follow [15] to search the giant graviton solution in the deformed spacetime.  We find that the configurations whose angular momentum $P$ are within a region  $P_L<P<P_U$ could have a fixed size and have lower energy than the point-like graviton.   The  lower (upper) critical angular momentum $P_L (P_U)$ is the decreasing (increasing) function of the electric/magnetic Melvin field.  The property is in contrast to that in the deformed $AdS_5 \times S^5$ spacetime.  In [14] we had showed that the giant graviton in the magnetic Melvin field deformed $AdS_5 \times S^5$ spacetime will become more stable than the point-like graviton if its angular momentum $P$ is smaller than a critical value $P_c$.  However,  the giant graviton in the electric Melvin field deformed $AdS_5 \times S^5$ spacetime is always unstable and will transit into a point-like graviton.   

  It is known that [20] the giant gravitons formed from a stack of  D1-branes in the near-horizon spacetime of D5-branes background have also the same special  properties as those in the  $AdS_3 \times S^3 \times T^4$, i.e. they exist when their angular momentum has a very specific value and, for this value of the momentum, the graviton can have arbitrary size.   In section IV we argue that the results found in section II and III could also be shown in the giant graviton in the near-horizon spacetime of  the electric/magnetic Melvin field deformed D5-branes background.   In section V we discuss our results. 

\section{Spinning String and Giant Graviton in Magnetic Melvin Field Deformed $AdS_3 \times S^3 \times T^4$}
\subsection{Magnetic Melvin Field Deformed $AdS_3 \times S^3 \times T^4$}
The metric of two D=11 M2-branes intersecting over a point  is  described by [17]
$$ds^2_{11}=-H_1^{-2\over3}H_2^{-2\over3} dt^2 + H_1^{-2\over3}H_2^{1\over3}\left(dx_1^2+dx_2^2\right)+H_1^{1\over3}H_2^{-2\over3}\left(dx_3^2+dx_4^2\right)~~~~~~$$
$$~~~~~~~~~~+H_1^{1\over3}H_2^{1\over3}\left (dz^2 + dx_{10}^2 + d\rho^2+\rho^2\left (d\theta^2 + \cos^2\theta d\phi^2 + \sin^2\theta d\chi^2\right)\right), \eqno{(2.1)}$$
where $x_1$, $x_2$ and $x_3$, $x_4$  are the internal coordinates of the two M2-branes.   The RR field strengths are 
$$F^{(4)} = - 3 dt \wedge dH_1^{-1} \wedge dx_1\wedge dx_2  - 3 dt \wedge dH_2^{-1} \wedge dx_3\wedge dx_4.\eqno{(2.2)}$$
$H_i$ is the harmonic function defined by 
$$ H_i = 1+ {L\over r^{D-p-3}}, ~~~~~~~r^2\equiv z^2+ \rho^2+x_{10}^2 , 
~~~~L \equiv {16\pi G_D\,T_p \,N_i\over D-p-3 },\eqno{(2.3)}$$
in which  $G_D$ is the D-dimensional Newton's constant, $T_p$ the p-brane tension, and $N_i$ the number of M2-brane. In the case of (2.1), $D=11$ and $p=2$. 

  We now transform the angle $\phi$ by mixing it with the compactified  coordinate $x_{10}$ in the following substituting 
$$\phi \rightarrow \phi + B x_{10},\eqno{(2.4)}$$ 
to obtain a magnetic Melvin flux  [16].  After the above substituting the line element can be expressed as 
$$ds_{11}^2= e^{-2\Phi/3}ds_{10}^2+  e^{4\Phi/3} (dx_{10}+2 A_\mu dx^\mu )^2 , \eqno{(2.5)} $$
In the above equation we have the relations
$$e^{4\Phi\over 3} = H_1^{1\over3}H_2^{1\over3}\sqrt{1+B^2\,\rho^2\cos^2\theta},\eqno{(2.6)}$$
$$A_\phi = {B\,\rho^2\cos^2\theta\over 2 \left(1+B^2\rho^2\cos^2\theta\right)},\hspace{1cm}\eqno{(2.7)}$$
in which $\Phi$ is the corresponding dilaton field and $A_\phi$ is called as a magnetic Melvin field [16].  Note that although the $B$ in (2.4) is a constant value the magnetic field strengths $F_{\phi\rho}$ and $F_{\phi\theta}$ calculated from (2.7) are non-uniform. 

   After the Kaluza-Klein reduction the 10D line element becomes
$$ds^2_{10}=\sqrt {1+ B^2\rho^2\cos^2\theta}\left[ -H_1^{-1\over2}H_2^{-1\over2} dt^2 + H_1^{-1\over2}H_2^{1\over2}\left(dx_1^2+dx_2^2\right)+H_1^{1\over2}H_2^{-1\over2}\left(dx_3^2+dx_4^2\right) \right.\hspace{3cm}$$
$$\left.\hspace{1.8cm}+H_1^{1\over2}H_2^{1\over2}\left (dz^2 + d\rho^2+\rho^2\left (d\theta^2 + {\cos^2\theta d\phi^2 \over 1+ B^2\rho^2\cos^2\theta} + \sin^2\theta d\chi^2\right)\right) \right],\eqno{(2.8)}$$
which describes the metric of a stack of  D2-branes  $\bot$ D2-branes with magnetic Melvin flux  in 10 D IIA string theory.  In this decomposition into ten-dimensional fields which do not depend on the $x_{10}$ the ten-dimensional Lagrangian density becomes 
$$ {\cal L}_{10D} = {\cal R}- 2 (\nabla \Phi)^2 - e^{2\sqrt 3 \Phi}~ F_{\mu\nu}F^{\mu\nu},\eqno{(2.9)}$$
in which $F_{\mu\nu}$ is the EM field strength.

  To find the corresponding background  of a stack of  D1-branes  $\bot$ D5-branes we will perform a series of the T-duality transformation [18] on the coordinates $z,~x_3,~x_4$. Using the formula that the metric and dilation field are replaced by 
$$g_{x_i x_i}\rightarrow {1\over g_{x_i x_i}}, ~~~~~~e^{\Phi}\rightarrow {e^{\Phi}\over \sqrt{g_{x_i x_i}}},\eqno{(2.10)}$$
in the case of T-duality transformation on the coordinates $x_i$, the background of  the magnetic Melvin field deformed system of a stack of D1-branes  $\bot$ D5-branes  becomes
$$ds^2_{10}=\sqrt {1+ B^2\rho^2\cos^2\theta}\left[H_1^{-1\over2}H_2^{-1\over2}\left(-dt^2+{dz^2\over 1+ B^2\rho^2\cos^2\theta}\right) + H_1^{1\over2}H_2^{-1\over2}\left(dx_1^2+dx_2^2 \hspace{2cm}\right.\right.$$
$$ ~~~~\left.\left. + {dx_3^2+dx_4^2\over 1+B^2\rho^2\cos^2\theta }\right) +H_1^{1\over2}H_2^{1\over2}\left (d\rho^2+\rho^2\left (d\theta^2 + {\cos^2\theta d\phi^2\over 1+ B^2\rho^2\cos^2\theta} + \sin^2\theta d\chi^2\right)\right)\right],\eqno{(2.11)}$$
in which
$$H_1=1+{Q_1\over \rho^2},~~~~H_2=1+{Q_5\over \rho^2},\eqno{(2.12)}$$
and RR field strengths are 
$$F_{tz\rho}= \partial_\rho H_1^{-1}, ~~~~~~F_{\alpha\beta\gamma}= -\rho^3\partial_\rho H_5^{-1}\sqrt {g_{S_{def}^3}}.\eqno{(2.13)}$$
The $Q_1$ and $Q_5$ in (2.12) denote the D1 and D5 charges in a proper unit,  as could be read from (2.3).  The $\alpha,~\beta,~\gamma$ are the coordinates $\theta$, $\phi$, $\chi $ of deformed $S^3$ (denoted as $S_{def}^3$) and the $g_{S_{def}^3}$is the determinant of the corresponding  metric as shown in (2.11). 

  In the near-horizon limit $\rho \rightarrow 0$ we can approximate $H_1\rightarrow {Q_1\over \rho^2}$, $H_2\rightarrow{Q_5\over \rho^2}$ and  line element (2.11) becomes
$$ds^2_{10}=\sqrt {1+ B^2\rho^2\cos^2\theta}\left[ {\rho^2\over \sqrt {Q_1Q_5}}\left(-dt^2+{dz^2\over 1+ B^2\rho^2\cos^2\theta}\right) + {Q_1\over Q_5} \left(dx_1^2+dx_2^2 \hspace{2cm}\right.\right.$$
$$~~~~~\left.\left. + {dx_3^2+dx_4^2\over B^2\rho^2\cos^2\theta }\right) +{\sqrt {Q_1Q_5}\over \rho^2}d\rho^2+\sqrt {Q_1Q_5}\left (d\theta^2 + {\cos^2\theta d\phi^2 \over 1+ B^2\rho^2\cos^2\theta} + \sin^2\theta d\chi^2\right) \right],\eqno{(2.14)}$$
\\
and  RR field strengths become
$$F_{tz\rho}= {2 \rho\over Q_1}, ~~~~~~F_{\alpha\beta\gamma}= 2Q_5 \sqrt {g_{S_{def}^3}}.\eqno{(2.15)}$$
In the case of  $B=0$  the above spacetime becomes the well-known geometry of $AdS_3\times S^3 \times T^4$. Thus, this background describes the magnetic Melvin field deformed $AdS_3\times S^3 \times T^4$.  

%%%%%%%%%%%%%%%%%%%%%%%%%%%%
\subsection{Spinning String  in Magnetic Melvin Spacetime}
We will follow [3] to search the string solution which is fixed on the spatial coordinates in deformed $S^3$ and locating at  $x_i=z=0$, $\theta=\theta_0$ in the magnetic field deformed spacetime (2.14) with a fixed value of $\rho=1$.  The line element becomes
$$ds_3^2 = \sqrt {1+ B^2cos^2\theta_0}\left[-dt^2  + {\cos^2\theta_0 d\phi^2 \over 1+ B^2\cos^2\theta_0} + \sin^2\theta_0 d\chi^2\right] ,\eqno{(2.16)}$$ 
in unit of $Q_1=Q_5=1$.  The string action could now be written in the conformal gauge in terms of the independent  global coordinates $x^m$  
$$ I= - { 1 \over 4 \pi} \int d^2 \xi \  G_{mn}(x) \partial_a x^m \partial^a  x^n, \eqno{(2.17)}$$
in which $\xi^a=(\tau,\sigma)$ and we let $\alpha' = 1$ for convenience.  In the conformal gauge $\sqrt {-g} g^{ab} = \eta^{ab}= $diag(-1,1), the equations of motion following from the action should be supplemented by the conformal gauge  constraints
$$ G_{mn}(x) ( \dot  x^m \dot  x^n +  x'^m   x'^n) =0 , \eqno{(2.18a)}$$
$$ G_{mn}(x) \dot x^m   x'^n =0.  \eqno{(2.18b)}$$
Following the method of Frolov and Tseytlin [3] we now adopt  the ansatz 
$$ t =\kappa \tau,~~~ \chi = \chi(\sigma),~~~\phi = \omega \tau.  \eqno{(2.19)}$$
to find the rotating string solution. 

Substituting the ansatz (2.19) into metric form (2.17) the associated Lagrangian  is 
$$L = - {1\over 4 \pi}\left[\sqrt {1+ B^2cos^2\theta_0}~\kappa^2  - {\omega^2\cos^2\theta_0  \over \sqrt{1+ B^2\cos^2\theta_0}} + \sqrt {1+ B^2cos^2\theta_0}\sin^2\theta_0 ~\chi(\sigma)'^2 \right] .\eqno{(2.20)}$$
As the deformation we used does not change the properties of the  translational isometries of  coordinates $t$, $\phi$ there are the corresponding two integrals of motion:
$$ {\cal E}= P_{t}= \int^{2\pi}_0 {d\sigma\over 2\pi}\sqrt {1+ B^2cos^2\theta_0}\,\partial _0 t \,,  \eqno{(2.21)}$$
which is the energy of the solution, and  
$$ J= P_{\phi}= \int^{2\pi}_0{d\sigma\over 2\pi}  {\cos^2\theta_0  \over \sqrt{1+ B^2\cos^2\theta_0}}~\partial _0 \phi \eqno{(2.22)}$$
which is the angular momentum of the rotating string in the magnetic field deformed $S^3$ space. 

 To find the values of energy and angular momentum we must know the function of $\chi(\sigma)$ and the relation between $\kappa$ and $\omega$.  This can be obtained by solving the equations of  $\chi(\sigma)$ associated to the Lagrangian (2.20), and imposing the conformal gauge constraints of (2.18).  The field equation of $\chi(\sigma)$ is
$$0= \left(\sqrt {1+ B^2\cos^2\theta_0}\sin^2\theta_0 ~\chi(\sigma)'\right)',\eqno{(2.23)}$$
which could be easily solved by setting
$$ \chi(\sigma) = n \sigma, \eqno{(2.24)}$$
which are the same as those in the undeformed space.  Using the above relation 
we see that while the conformal gauge constraints (2.18b) is automatically satisfied the another conformal gauge constraints of (2.18a) implies
$$\left(1+ B^2cos^2\theta_0\right)\left(\kappa^2 - n^2 (1- cos^2\theta_0)\right) - \omega^2cos^2\theta_0 = 0.\eqno{(2.25)}$$
Using the above relations the energy and momentum of the string solutions have the simple forms 
$${\cal E}(\cos\theta_0) =\sqrt{ \omega^2\cos^2\theta_0 + n^2 (1+ B^2\cos^2\theta_0)\sin^2\theta_0 }.\eqno{(2.26a)}$$
$$J(\cos\theta_0) = {\cos^2\theta_0\over \sqrt{1+ B^2 \cos^2\theta_0}}~\omega.\hspace{3.5cm}\eqno{(2.26b)}$$
Eq.(2.26a)  implies that 
$${d{\cal E}\over d \theta_0} = {\left(n^2(1+B^2) - \omega^2\right)\sin\theta_0~\cos\theta_0- 2 n^2B^2 \sin^3\theta_0~\cos\theta_0\over \sqrt{\omega^2 + \left(n^2(1+B^2) - \omega^2\right) \sin^2\theta_0-n^2B^2\sin^4\theta_0}},\eqno{(2.26c)}$$
and solutions of  $d{\cal E}/ d \theta_0=0$ could be shown at $\sin\theta_0 = 0$ and $\sin\theta_0 = 1$ in which the corresponding string energy are ${\cal E}(\sin\theta_0=0) =\omega $ and ${\cal E}(\sin\theta_0=1) =n $ respectively.  (We can also from the relation ${\cal E}^2 = \omega^2 + \left(n^2(1+B^2) - \omega^2\right) \sin^2\theta_0-n^2B^2\sin^4\theta_0$  see that the energy is minimum at $\sin\theta_0=1$ if $n < \omega$ and is minimum at $\sin\theta_0=0$  if  $\omega < n$.)

  Therefore we have two configurations of least energy.   The first is 
$${\cal E}(\sin\theta_0=1) = n,~~~~~J(\sin\theta_0=1) = 0, ~~~~~{if}~~n < \omega .\eqno{(2.27)}$$
In this case $\cos\theta_0=0$ and from (2.20) we can see that $\phi$ decouples from the Lagrangian and there is not any time evolution of the string along this coordinate.  The solution is static and has a zero angular momentum, which is not a desired one as the AdS/CFT duality in here [2-6] is comparing the rapidly spinning string with the corresponding anomalous dimension in gauge theory.

 The next useful configuration is
$${\cal E}(\sin\theta_0=0) = \omega,~~~~~J(\sin\theta_0=0) =  {\omega\over \sqrt{1+ B^2}}, ~~~~~{if}~~\omega < n,\eqno{(2.28a)}$$
which implies that 
$${\cal E} =  {J~\sqrt{1+ B^2}}> J.\eqno{(2.28b)}$$
Thus the external electric flux will increase the string energy and, from the AdS/CFT point of view, the correction of the anomalous dimensions of operators in the dual field theory will be positive [3-6].   Note that the solution is at $\sin\theta_0 = 0$ and we have a string shrunk to a point and this point is circling around $\phi$ cycle.   This means that, in contrast to the string solutions found in previous literatures [3-5,13] in which the closed string solution has a finite radius,  the above configuration is a point-like string spinning along the $\phi$ coordinate with an angular momentum $J$.  
%%%%%%%%%%%%%%%%%%%%%%%%%%%%
\subsection{Giant Graviton in Magnetic Melvin Spacetime}
The ``genuine giant graviton'' found first in [7] is the rotating D3-brane wrapping the  spherical part spacetime of $S^5$ in the $AdS_5 \times S^5$ background.  Later, the ``dual giant graviton'' found in [8] is the rotating D3-brane wrapping the spherical part spacetime of $AdS_5$ in the $AdS_5 \times S^5$ background.   The genuine giant graviton has zero size in the $AdS_5$ and dual giant graviton has zero size in the $S^5$.  In this section, however, we will consider only the ``genuine giant graviton'' expanding in the deformed $S^3$.   This is because that the corresponding deformed $AdS_3$ spacetime in (2.14)  is mathematically complex and it is difficult to study the behavior of how a brane wrapping on it.

  Therefore, letting  $\sigma_0$, $\sigma_1$ be the worldsheet coordinates of the giant graviton which is a bound state of $n$ D1-branes and $m$ D5-branes wrapped on the deformed 4-torus \footnote[1]{The coordinates of the deformed 4-torus are $x_1$, $x_2$, $x_3$ and $x_4$ as shown on (2.14).  As the D5-branes are wrapping on the deformed 4-torus there remains only two worldsheet coordinates.   Therefore the {\it macroscopic} giant graviton has only the worldsheet coordinates $\sigma_0$ and $\sigma_1$.},  we will follow [15] to look at a trial solution with the ansatz
$$\sigma_0= t,~~\sigma_1 = \chi,~~~~\rho= constant,~~~\theta=constant,~~~z=0,~~~~~\phi=\phi(t),\eqno{(2.29)}$$
which describes a giant graviton running around the sphere along the coordinate $\phi$.    In this case the 2-form RR potential on the deformed $S^3$ calculated from (2.15) is 
$$A^{(2)}_{\phi\chi} = Q_5\sin^2\theta\left(1+ {2\over B^2\rho^2} -{2\over B^2\rho^2} \sqrt {1+B^2\rho^2 \cos^2\theta}\right).\eqno{(2.30)}$$
Note that $A^{(2)}_{\phi\chi}\rightarrow Q_5 \sin^2\theta$ as $B \rightarrow 0$, which is that used in the undeformed theory [15].

The action of a bound state of $n$ D1-branes and $m$ D5-branes D1-brane can be written as
$$S=    n T_1 \left(- \int d^{2}\sigma\ \sqrt{-g} + \int P[A^{(2)}]\right) +  m T_5 \left(- \int  d^{4}\sigma\ \sqrt{-g}+  \int P[A^{(4)}]\right),\eqno{(2.31)}$$
in which $T_1$ is the D1-brane tension and $T_5$ the D5-brane tension.  $g_{ij}$ is the pull-back of the spacetime metric to the world-volume, and $P[A^{(i)}]$ denotes the pull-back of the $i$-form potential. 

  Now we can use the classical rotating D3-brane solution ansatz  (2.29) to find the associated Lagrangian.  After the calculations the action of the giant graviton becomes
$$S = -\left(2\pi n T_1 + (2\pi)^5 m T_5 {Q_1\over Q_5}\right)\sqrt{Q_1Q_5} \int dt \left[\sin\theta \sqrt{\rho^2 (1+B^2\rho^2\cos^2\theta) -Q_1Q_5 \cos^2\theta~\dot\phi^2}\right. $$
$$\left. - Q_5\sin^2\theta\left(1+ {2\over B^2\rho^2} -{2\over B^2\rho^2} \sqrt{1+B^2\rho^2 \cos^2\theta}\right)~\dot\phi \right], \eqno{(2.32)}$$
In the limit of $B\rightarrow 0$ above action becomes that in undeformed system [15] . Note that we have compactified the deformed 4-torus coordinate by choosing $0\le x_i\le2\pi$.

   To proceed, we will choose the special  units of the energy H, angular momentum $P_\phi $, and magnetic flux $B$, then, after the calculations, the angular momentum and energy of the deformed giant graviton have the relations 
$$P\equiv P_\phi  = {\partial {\cal L}\over \partial \dot\phi}
=R^2 \left(1+ {2\over B^2} -{2\over B^2} \sqrt{1+B^2 (1-R^2)}\right)  + {R(1-R^2) \dot\phi  \over \sqrt{1+B^2 (1-R^2) -(1-R^2)\dot\phi^2 }} \eqno{(2.33)}$$
 
$$ H =P\dot\phi- {\cal L} =  {\sqrt{1+B^2\,(1-R^2)}\over \sqrt{1-R^2}}\sqrt{\left[P -\left(1+ {2\over B^2} -{2\over B^2} \sqrt{1+B^2 (1-R^2)}\right)\right]^2+ R^2(1-R^2)},\eqno{(2.34)}$$
in which 
$$R \equiv \sin\theta, \eqno{(2.35)}$$
is the radius of the giant graviton in our notation.  We now use (2.34) to analyze the solution of giant graviton.  First, in the limit of undeformed system,  $B \rightarrow 0$, (2.34) becomes
$$H \approx \sqrt{P^2 + R^2 - 2PR^2\over 1-R^2}
-B^2 { (P-2P^2-3R^2+3PR^2+R^4)~\sqrt {1-R^2}~\over  \sqrt{P^2 + R^2 - 2PR^2}}+ O(B^4). \eqno{(2.36)}$$
In the case of $B=0$ the energy in (2.36) is an increasing function of $R^2$ and there is not any giant graviton solution in general.  However, the configuration with a special case of  $P=1$ will have a constant energy $H=1$, which is independent of the size of the giant graviton.   Thus,  undeformed giant graviton only exists when its angular momentum is a specific value and, moreover, it could have arbitrary size.  This property was already noted in the initial paper [7].

   Next , to see the effect of the magnetic Melvin field on the  giant graviton  we first use the formula (2.34) to plot the energy  H of the deformed giant graviton with angular momentum $P=0.6$ as a function of its radius R under a fixed magnetic flux $B=2$ and $B=6$ in figure 1.   The figure explicitly shows that the giant graviton has lower energy than the trivial graviton which becomes a metastable state and will tunnel into the nontrivial configuration of giant graviton.   The stable giant graviton has a finite radius at $R \approx 0.89$ if $B=2$ and  $R \approx 0.92$ if $B=6$.  The numerical analyses have shown a general property that {\it \bf the giant graviton size is an increasing function of the Melvin field}.  
\\

\hfil\scalebox{1}{\includegraphics{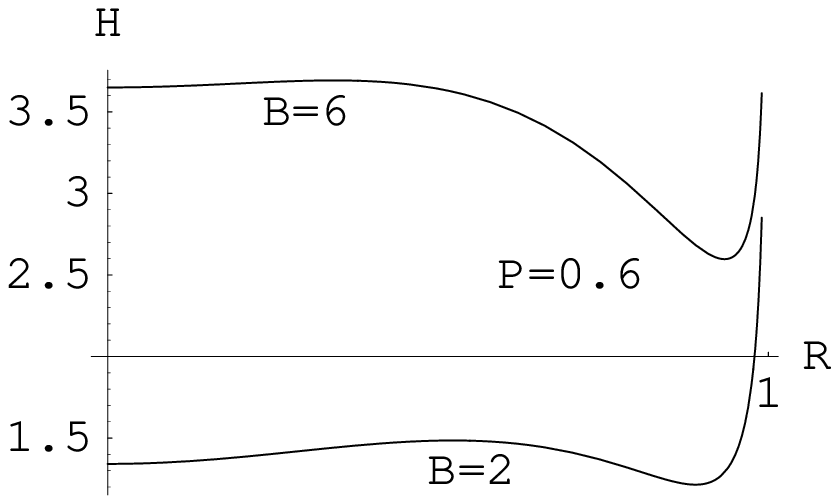}}\hfil\\
\\
{\it Figure 1:  Energy H of a  giant graviton with angular momentum $P= 0.6$ as the function of its radius R under a magnetic flux $B=2$ and $B=6$.  The point-like trivial graviton becomes a metastable state and will tunnel into the nontrivial configuration of giant graviton. It is seen that the giant graviton size is an increasing function of the Melvin field}
\\

   We also use the formula (2.34) to plot the energy H of the deformed giant graviton as a function of its radius $R$ with various angular momentum $P$ under a fixed magnetic flux $B=2$.   The results are shown in figure 2.   We see that {\it \bf while increasing the angular momentum of giant graviton the radius of the giant graviton is initially an increasing function}, the property is the same as that in the ``genuine giant graviton'' found in [7].   However, {\it \bf after the radius of the giant graviton reaches its maximum value  $R=1$ at $P=1$ it then becomes a decreasing function of the angular momentum}.   In these regions the giant graviton is still a stable configuration, in contrast to the ``genuine giant graviton'' in [7]. 
\\

\hfil\scalebox{1}{\includegraphics{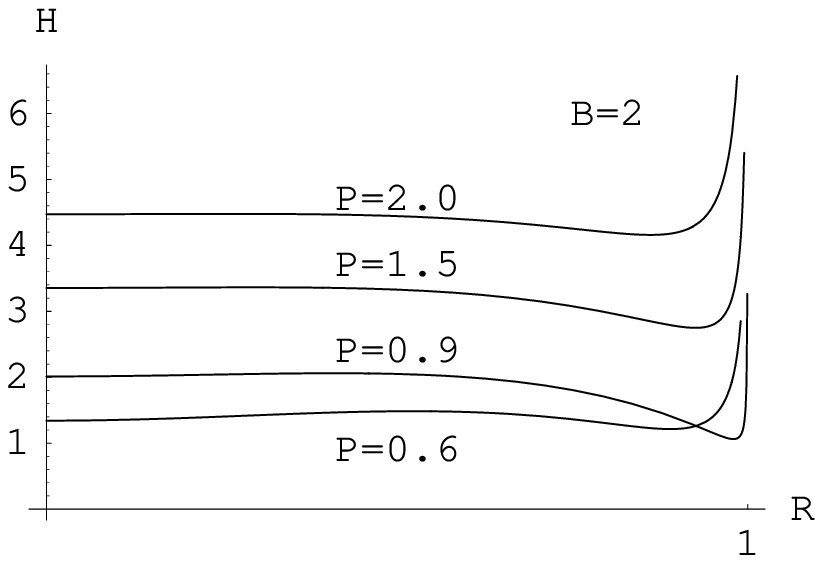}}\hfil\\
\\
{\it Figure 2:  Energy  H of the giant graviton with various angular momentum P=0.6, 0.9, 1.5 or 2 as the function of its radius $R$ under a magnetic flux $B=2$. Giant graviton has maximum radius  $R=1$ as $P=1$.}
\\

   We have performed many numerical analyses and see that, {\it \bf under a fixed Melvin flux only the configurations whose angular momenta $P$ are within a finite region, $P_L < P < P_U$, could have a fixed size and have lower energy than the point-like graviton.   The  lower (upper) critical angular momentum $P_L (P_U)$ is the decreasing (increasing) function of the Melvin field}.  In figure 3 we plot the phase diagram of the magnetic field deformed giant graviton and, for example, under the magnetic Melvin flux $B=2$, it is seen that the stable giant graviton with a finite radius could exist only if its angular momentum satisfies the relation $ P_L = 0.56 < P < 3.14 = P_U$.   
\\

\hfil\scalebox{1}{\includegraphics{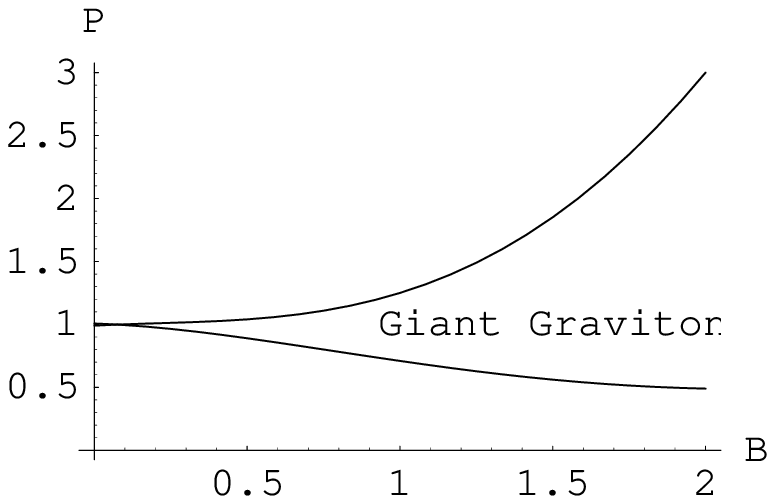}}\hfil\\
\\
{\it Figure 3:  Phase diagram of the magnetic field deformed giant graviton.  A stable giant graviton could appear only if its angular momentum $P$ is within a finite region $P_L < P < P_U$.}
\\

  It shall be noted that the above property is in contrast to that of the giant graviton in the deformed $AdS_5 \times S^5$, which was studied in our previous paper [14].  In [14] we had found that an arbitrary value of the Melvin flux could make the giant graviton more stable than the trivial graviton if its angular momentum $P$ is smaller than a critical value $P_c$. 

  An usefully analytic result may be obtained from (2.34) in the case of  $P \approx 1$ under a small value of $B$.  In this case, the giant graviton has a finite radius $R_{gg} \rightarrow 1$.  The energy of the giant graviton $H(R_{gg})$  and trivial graviton $H(0)$ are 
$$H(R_{gg}) \approx  P + {1\over 4}B^2 (1-P)^2, ~~~~~~H(0)  \approx  P + {1\over 2}B^2 P^2, \eqno{(2.37)}$$
respectively.   Thus
$$H(R_{gg})- H(0)  \approx  -  {1\over 2}B^2 < 0, \eqno{(2.38)}$$
which means that the deformed giant graviton is more stable than the trivial graviton.  When $B=0$ then (2.37) implies the relation $H = P $ in agreement with the BPS bound of the giant graviton.  Also, under a Melvin flux we see that $H(R_{gg}) > P$ which means that the external magnetic flux will increase the string energy.  Therefore, from the AdS/CFT point of view, the corrections of the anomalous dimensions of operators in the dual field theory will be positive [9,10].  

 %%%%%%%%%%%%%%%%%%%%%%%%%%%
\section{Spinning String and Giant Graviton in Electric Melvin Field Deformed $AdS_3 \times S^3 \times T^4$}
\subsection{Electric Melvin Field Deformed $AdS_3 \times S^3 \times T^4$}
To find the electric Melvin field deformed $AdS_3 \times S^3 \times T^4$ we first transform the time by mixing it with the compactified coordinate $x_{10}$ in the following substituting [19]
   $$t \rightarrow t - E x_{10}.\eqno{(3.1)}$$ 
Using the above substitution the line element (2.1) could be expressed as the Kaluza-Klein metric (2.5) in which 
$$ds_{10}^2= \sqrt {1-E^2H_1^{-1}H_2^{-1}}\left[ -{H_1^{-1\over2}H_2^{-1\over2}\over 1-E^2H_1^{-1}H_2^{-1} }dt^2 + H_1^{-1\over2}H_2^{1\over2}\left(dx_1^2+dx_2^2\right)+H_1^{1\over2}H_2^{-1\over2}\left(dx_3^2+dx_4^2\right) \right.\hspace{3cm}$$
$$\left.\hspace{1.8cm}+H_1^{1\over2}H_2^{1\over2}\left (dz^2 + d\rho^2+\rho^2\left (d\theta^2 + \cos^2\theta d\phi^2 + \sin^2\theta d\chi^2\right)\right) \right], \eqno{(3.2)}$$ 
$$e^{4\Phi\over 3} = H_1^{1\over3}H_2^{1\over3}\left(1-E^2H_1^{-1}H_2^{-1}\right) ,~~~~\hspace{10cm}\eqno{(3.3)}$$
$$A_t = {H_1^{-1}H_2^{-1}\over 2 \left(1-E^2H_1^{-1}H_2^{-1}\right)}, \hspace{11.5cm}\eqno{(3.4)}$$ 
in which  $\Phi$ is the corresponding dilaton field and $A_t$ is called as an electric Melvin field [19].

  The corresponding background  of a stack of  D1-branes  $\bot$ D5-branes could be obtained by the T-duality transformations [18] on the coordinates $z,~x_3,~x_4$. Using the formula (2.10)  the background of  the electric Melvin field deformed system of a stack of D1-branes  $\bot$ D5-branes  becomes
$$ds_{10}^2= \sqrt {1-E^2H_1^{-1}H_2^{-1}}\left[{H_1^{-1\over2}H_2^{-1\over2}\over 1-E^2H_1^{-1}H_2^{-1}}\left(-dt^2+ dz^2\right) + H_1^{1\over2}H_2^{-1\over2}\left(dx_1^2+dx_2^2\right.\right.\hspace{2cm}$$
$$\left.\left.+{dx_3^2+dx_4^2\over 1-E^2H_1^{-1}H_2^{-1}}\right) +H_1^{1\over2}H_2^{1\over2}\left (d\rho^2+\rho^2\left (d\theta^2 + \cos^2\theta d\phi^2 + \sin^2\theta d\chi^2\right)\right) \right], \eqno{(3.5)}$$ 
In the near-horizon limit $\rho \rightarrow 0$,  we can approximate $H_1\rightarrow {Q_1\over \rho^2}$, $H_2\rightarrow{Q_5\over \rho^2}$ and the line element (3.5) becomes
$$ds^2_{10}=\sqrt {1- {E^2\rho^4\over Q_1Q_5}}\left[ {\rho^2\over \sqrt {Q_1Q_5}}{1\over 1- {E^2\rho^4\over Q_1Q_5}}\left(-dt^2+dz^2\right) + {Q_1\over Q_5} \left(dx_1^2+dx_2^2 + {dx_3^2+dx_4^2\over 1- {E^2\rho^4\over Q_1Q_5}}\right) \hspace{2cm}\right.$$
$$ \hspace{1.5cm}\left. +{\sqrt {Q_1Q_5}\over \rho^2}d\rho^2+\sqrt {Q_1Q_5}\left (d\theta^2 + \cos^2\theta d\phi^2 + \sin^2\theta d\chi^2\right) \right].\eqno{(3.6)}$$
The RR field strengths are as those described in (2.15).  In the case of  $E=0$  the above spacetime becomes the well-known geometry of $AdS_3\times S^3 \times T^4$. Thus, the background describes the electric Melvin field deformed $AdS_3\times S^3 \times T^4$.  
%%%%%%%%%%%%%%%%%%%%%%%%%%%%
\subsection{Spinning String  in Electric Melvin Spacetime}
We now search the string solution which is fixed on the spatial coordinates in deformed $S^3$ and locating at  $x_i=z=0$, $\theta=\theta_0$ in the electric field deformed spacetime (3.6) with a fixed value of $\rho=1$.  The line element therefore becomes
$$ds_3^2 = -{dt^2\over \sqrt {1- E^2}}  + \sqrt {1- E^2}\cos^2\theta_0 d\phi^2 - \sqrt {1- E^2}\sin^2\theta_0 d\chi^2 ,\eqno{(3.7)}$$ 
in unit of $Q_1=Q_5=1$.  Following the process in previous section we will now adopt  the ansatz (2.19) to find the rotating string solution.  The associated Lagrangian  is 
$$L = - {1\over 4 \pi}\left[{\kappa^2 \over \sqrt {1- E^2}} -\sqrt {1- E^2}~ \omega^2\cos^2\theta_0 + \sqrt {1- E^2} \sin^2\theta_0 ~\chi (\sigma)'^2\right].\eqno{(3.8)}$$

The corresponding energy and angular momentum of the rotating string in the electric field deformed $S^3$ space are
$$ {\cal E}= P_{t}= \int^{2\pi}_0 {d\sigma\over 2\pi}{1\over \sqrt {1- E^2}}\,\partial _0 t , ~~~~~~~~~ \eqno{(3.9)}$$ 
$$ J= P_{\phi}= \int^{2\pi}_0{d\sigma\over 2\pi}  \sqrt {1- E^2}\cos^2\theta_0  ~\partial _0 \phi , \eqno{(3.10)}$$
respectively.  The field equation of $\chi(\sigma)$ is
$$0= \left(\sqrt {1- E^2} \sin^2\theta_0 ~\chi (\sigma)'\right)',\eqno{(3.11)}$$
which could be easily solved by setting
$$ \chi(\sigma) = n \sigma, \eqno{(3.12)}$$
which are the same as those in the undeformed space.  Using the above relation 
we see that while the conformal gauge constraints (2.18b) is automatically satisfied the another conformal gauge constraints of (2.18a) implies
$$\kappa^2 =(1- E^2)\left[\omega^2 \cos^2\theta_0 + n^2 \sin^2\theta_0\right].\eqno{(3.13)}$$
Using the above relations the energy and momentum of the string have the simple forms 
$${\cal E}(\theta_0) =\sqrt{ \omega^2 \cos^2\theta_0 + n^2 \sin^2\theta_0}.\eqno{(3.14a)}$$
$$J(\theta_0) =\sqrt{1-E^2}~ \omega \cos^2\theta_0.~~~~~~\eqno{(3.14b)}$$
Eq.(3.14a)  implies that 
$${d{\cal E}\over d \theta_0} = {(n^2- \omega^2 ) \sin\theta_0~\cos\theta_0\over \sqrt{\omega^2 + (n^2 - \omega^2) \sin^2\theta_0}},\eqno{(3.14c)}$$
and solutions of  $d{\cal E}/ d \theta_0=0$ could be shown at $\sin\theta_0 = 0$ and $\sin\theta_0 = 1$ in which the corresponding string energy are ${\cal E}(\sin\theta_0=0) =\omega $ and ${\cal E}(\sin\theta_0=1) =n $ respectively.  (We can also from the relation ${\cal E}^2 = \omega^2 + (n^2 - \omega^2) \sin^2\theta_0$  see that the energy is minimum at $\sin\theta_0=1$ if $n < \omega$ and is minimum at $\sin\theta_0=0$  if  $\omega < n$.)

  Therefore we have two configurations of least energy.   The first is 
$${\cal E}(\sin\theta_0=1) = n,~~~~~J(\sin\theta_0=1) = 0, ~~~~~{if}~~n < \omega .\eqno{(3.15)}$$
In this case $\cos\theta_0=0$ and from (3.7) and (3.8) we can see that $\phi$ decouples from the Lagrangian and there is not any time evolution of the string along this coordinate.  The solution is static and has a zero angular momentum, which is not a desired one as the AdS/CFT duality in here [2-6] is comparing the rapidly spinning string with the corresponding anomalous dimension in gauge theory.

 The next useful configuration is
$${\cal E}(\sin\theta_0=0) = \omega,~~~~~J(\sin\theta_0=0) = \omega \sqrt{1-E^2}, ~~~~~{if}~~\omega < n,\eqno{(3.16a)}$$
which implies that 
$${\cal E} =  {J\over \sqrt{1- E^2}}> J.\eqno{(3.16b)}$$
Thus the external electric flux will increase the string energy and, from the AdS/CFT point of view, the correction of the anomalous dimensions of operators in the dual field theory will be positive [3-6].   Note that the solution is at $\sin\theta_0 = 0$ and we have a string shrunk to a point and this point is circling around $\phi$ cycle.   This means that, in contrast to the string solutions found in previous literatures [3-5,13] in which the closed string solution has a finite radius,  the above configuration is a point-like string spinning along the $\phi$ coordinate with an angular momentum $J$.  
%%%%%%%%%%%%%%%%%%%%%%%%%%%%
\subsection{Giant Graviton in Electric Melvin Spacetime}
We now consider a giant graviton configuration as that described in equation (2.29).   Through the standard procedure as in previous section we can obtain the action of the giant graviton 
$$S = -\left[2\pi n T_1 + (2\pi)^5 m T_5 {Q_1\over Q_5}\right]\sqrt{Q_1Q_5} \int dt \left[\sin\theta \sqrt{{\rho^2\over \sqrt{Q_1Q_5}}  -\left(1- {E^2\rho^4\over Q_1Q_5}\right) Q_1Q_5 \cos^2\theta~\dot\phi^2}\right.$$
$$\left.\hspace{10cm}- Q_5 \sin^2\theta~\dot\phi \right]. \eqno{(3.17)}$$
Choose the units as before and after the calculations the associated angular momentum and energy of the deformed giant graviton have the relations
$$P\equiv P_\phi  = {\partial {\cal L}\over \partial \dot\phi}
=R^2  +  {{1-E^2}~R(1-R^2) \dot\phi  \over \sqrt{1-(1-E^2)(1-R^2)\dot\phi^2 }}, \hspace{2cm}\eqno{(3.18)}$$
 $$ H =P\dot\phi- {\cal L} ={(P-R^2)^2 + \sqrt{1-E^2}\,R^2 (1-R^2)\over \sqrt{1-E^2} \sqrt{1-R^2} \sqrt{(P-R^2)^2+ R^2 (1-R^2)}} ,\eqno{(3.19)}$$
in which $R \equiv \sin\theta$ is the radius of the giant graviton in our notation. 

  We can now use (3.19) to plot the energy of the deformed giant graviton as a function of its radius $R$ with various angular momentum $P$ under a fixed electric flux.     With the help of  more numerical analyses we have seen that {\it \bf the effects of the electric Melvin field on the giant graviton are the same as those of the magnetic Melvin field}.  Therefore the conclusions drawn in the section II could also be use in the electric Melvin field background.  In figure 4 we plot the phase diagram of the electric field deformed giant graviton and, for example,  the lower critical angular momentum for the case of $E \rightarrow 1$ is found to be $P_L = 0$ and the upper one is $P_U  \approx 2.4$.
\\

\hfil\scalebox{1}{\includegraphics{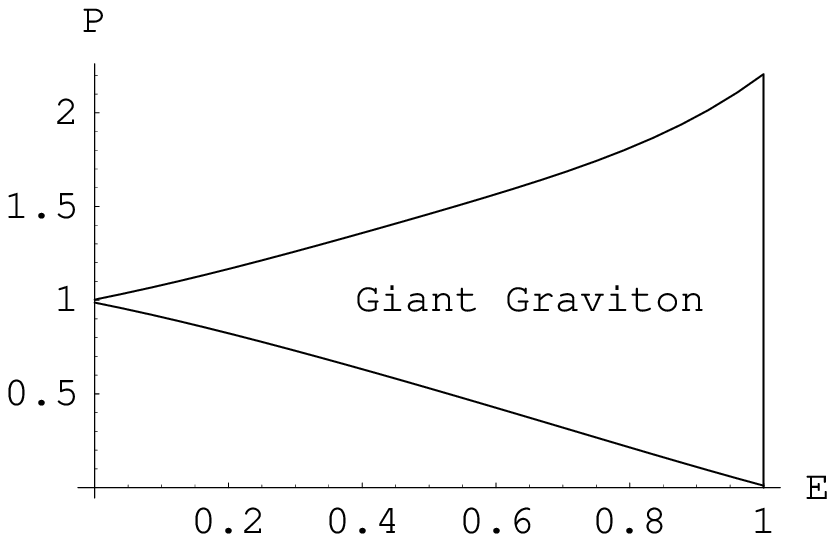}}\hfil\\
\\
{\it Figure 4:  Phase diagram of the electric field deformed giant graviton.  A stable giant graviton could appear only if its angular momentum $P$ is within a finite region $ P_L<P<P_U$.}
\\

 It is worthy to note that the giant graviton in the electric Melvin field deformed $AdS_5 \times S^5$ spacetime is always unstable and will transit into a point-like graviton, irrespective of its angular momentum [14].  

  An usefully analytic result may be obtained from (3.19) in the case of  $P \approx 1$ under a small value of $E$.  In this case, the giant graviton has a finite radius $R_{gg} \rightarrow 1$.  The energy of the giant graviton $H(R_{gg})$  and trivial graviton $H(0)$ are  
$$H(R_{gg}) \approx ~P +{|1-P|\over \sqrt{1-E^2}}, ~~~~~~H(0) \approx P~\left(1+{E^2\over2}\right), \eqno{(3.20)}$$
respectively.   Thus
$$H(R_{gg})- H(0)  \approx  -{E^2}  < 0 ,\eqno{(3.21)}$$
which means that the deformed giant graviton is more stable than the trivial graviton.  When $E=0$ then (3.20) implies the relation $H = P$  in agreement with the BPS bound of the giant graviton.  Also, under a electric Melvin field flux we see that $H(R_{gg}) > P$ which means that the external electric flux will increase the string energy.  Therefore, from the AdS/CFT point of view, the corrections of the anomalous dimensions of operators in the dual field theory will be positive [9,10].  

%%%%%%%%%%%%%%%%%%%%%%%%%%%%%
\section{Giant Graviton in  Melvin Field Deformed $D5$ Background}
   The giant gravitons formed from a stack of  D1-branes in the near-horizon spacetime of D5-branes background had been investigated in [20].   These authors had found that the giant gravitons could exist only when their angular momentum has a very specific value and, for this value of the momentum, the graviton can have arbitrary size,  the special  properties as those in the  $AdS_3 \times S^3 \times T^4$.   In this section we will argue that the results found in section II and III could also be shown in the giant graviton in the near-horizon spacetime of Melvin field deformed D5-branes background.   

   First, the magnetic/electric Melvin field deformed $D5$ background could be obtained from (2.11) and (3.5) after letting the $D1$ charge number described in (2.12) to be $Q_1=0$.   In this case the harmonic function described in (2.12) becomes $H_1 =1$. Second, in taking the near-horizon limit we have the relations $H_2 \rightarrow {Q_5\over \rho^2}$ and $H_1 =1$.  However, in taking the near-horizon limit in the $D1 \bot D5 $  background  we have  the relation  $H_1 \rightarrow {Q_1\over \rho^2}$. 

   Therefore, in near-horizon limit of the D5 background the metric will be just these in (2.14) and (3.6) with replacement of 
$${Q_1\over \rho^2} \rightarrow 1,  ~~~~~~~as~~~~~~ (D1~branes~\bot~D5~branes) \rightarrow (D5). \eqno{(4.1)}$$ 
However, the constants $Q_1$ and $\rho$ are irreverent to our analyses as they could be neglected after choosing  the proper units of the energy H, angular momentum $P$, and magnetic flux $B$, as discussed in sections II and III.  Therefore the background (2.14) and (3.6) could also be applied to analyze the spinning string and giant graviton in the magnetic/electric Melvin field deformed $D5$ background, in near-horizon limit. The properties of  the solutions will be like those in sections II and III.

    Note that, the ``dual giant graviton'' expanding in the deformed $AdS_3$ part of the space  will feel the character of the geometry.  As the near horizon geometry of the D5 background has not $AdS$ part [20],  the corresponding ``dual giant gravitons'' (if they exist) in the $D5$ background may have different behaviors from those in the $D1 \bot D5$  background.  Therefore, the above conclusion could at most be used for the ``genuine giant graviton'' expanding in the deformed $S^3$, which are investigated in previous sections. 

%%%%%%%%%%%%%%%%%%%%%%%%
\section{Conclusion}
In a previous work [14] we had investigated the giant graviton in the electric/magnetic Melvin field deformed $AdS_5 \times S^5$ spacetime.   As the giant graviton in the undeformed $AdS_5 \times S^5$ spacetime has the same energy as the point-like graviton the main result of [14] is to see that the magnetic Melvin field has an effect to stabilize the giant graviton and to suppress it from tunneling into the point-like graviton.  The electric Melvin field,  however, always renders the giant graviton unstable. 

In this paper we investigate the giant graviton in the electric/magnetic Melvin field deformed $AdS_3 \times S^3 \times T^4$.  The giant graviton in the undeformed $AdS_3 \times S^5 \times T^4$ spacetime has a special property that it only exists when its angular momentum is a specific value and, moreover, it could have arbitrary size [7,15].   Our investigations of this paper have found that the configurations whose angular momentum $P$ are within a region, $P_L<P<P_U$, could have a fixed size and have lower energy than the point-like graviton.   The lower (upper) critical angular momentum $P_L (P_U)$ is found to be the decreasing (increasing) function of the Melvin field.  The property is in contrast to that in the deformed $AdS_5 \times S^5$ spacetime.  The previous analyses in [14] showed that the giant graviton in the magnetic Melvin field deformed $AdS_5 \times S^5$ spacetime will become more stable than the point-like graviton if its angular momentum $P$ is smaller than a critical value $P_c$.  While the giant graviton in the electric Melvin field deformed $AdS_5 \times S^5$ spacetime is always unstable and will transit into a point-like graviton.   We also adopt an ansatz to find the classical string solutions which are rotating in the deformed $S^3$ with an angular momenta in the rotation plane.  We obtain the relations between the energy and its angular momentum for the  spinning string and giant graviton solutions and find that the external magnetic/electric flux will increase the solution energy.  Therefore, from the AdS/CFT point of view, the corrections of the anomalous dimensions of operators in the dual field theory will be positive [3-6,9,10].  

It is known that the supersymmetry is broken by the magnetic or electric field,  the spinning string and giant graviton in the electric/magnetic Melvin geometries could provide us the systems to investigate the non-supersymmetric examples of the gauge/gravity correspondence.  As the string theory in $AdS_{3}\times S^{3}$ is  dual to the $1+1$ CFT [1,10],  our investigations of the string and giant graviton in the electric/magnetic Melvin deformed $AdS_{3}\times S^{3}$ therefore are of interesting.  Finally, it is noted that several prospects of  the giant gravitons in the $AdS_3 \times S^3 \times T^4$ spacetime which are the correct supergravity description of the chiral primary states of the D1-D5 system could be described by adding the rotation [21].  Therefore, it is interesting to investigate the giant gravitons in the background of the rotating intersecting D-branes [22] under the Melvin field deformation.  The geometry in there is more complex and problem remains to be studied.
%%%%%%%%%%%%%%%%%%%%%%%
~
\\
~
\\
~
\\
~
{\bf  \large REFERENCES}
\begin{enumerate}
\item J.~M.~Maldacena, ``The large N limit of superconformal field theories and supergravity,'' Adv.\ Theor.\ Math.\ Phys.\  {\bf 2}, 231 (1998) [Int.\ J.\ Theor.\ Phys.\  38 (1999) 1113  [hep-th/9711200];  E.~Witten, ``Anti-de Sitter space and holography,'' Adv.\ Theor.\ Math.\ Phys.\   2 (1998) 253 [hep-th/9802150].
\item S.~S.~Gubser, I.~R.~Klebanov and A.~M.~Polyakov, ``Gauge theory correlators from non-critical string theory,'' Phys.\ Lett.\ B428 (1998) 105 [hep-th/9802109].
\item S.~Frolov and A.~A.~Tseytlin, ``Multi-spin string solutions in
$AdS_5 \times S^5$,'' Nucl.\ Phys.\ B668 (2003) 77 [hep-th/0304255]; S.~Frolov and A.~A.~Tseytlin, ``Quantizing three-spin string solution in $AdS_5 \times S^5$,'' JHEP 0307 (2003) 016 [hep-th/0306130].
\item S.~Frolov and A.~A.~Tseytlin, ``Rotating string solutions: AdS/CFT duality in non-supersymmetric sectors,'' Phys.\ Lett.\ B570 (2003) 96 [hep-th/0306143]; S.A. Frolov, I.Y. Park, A.A. Tseytlin, ``On one-loop correction to energy of spinning strings in $S^5$,''Phys.Rev. D71 (2005) 026006 [hep-th/0408187]; I.Y. Park, A. Tirziu, A.A. Tseytlin, ``Spinning strings in $AdS_5 \times  S^5$: one-loop  correction to energy in SL(2) sector,'' JHEP 0503 (2005) 013 [hep-th/0501203].
\item  G.~Arutyunov, S.~Frolov, J.~Russo and A.~A.~Tseytlin, ``Spinning strings in $AdS_5 \times S^5$ and integrable systems,'' Nucl.\ Phys.\ B671 (2003) 3 [hep-th/0307191]; G.~Arutyunov, J.~Russo and A.~A.~Tseytlin, ``Spinning strings in $AdS_5 \times S^5$: New integrable system relations,'' Phys.\ Rev.\ D69 (2004) 086009 [hep-th/0311004].
\item A.~A.~Tseytlin, ``Spinning strings and AdS/CFT duality,'' [hep-th/0311139]; J, Plefka, ``Spinning strings and integrable spin chains in the AdS/CFT correspondence,'' [hep-th/0507136].
\item  J. McGreevy, L. Susskind and N. Toumbas, ``Invasion of Giant Gravitons from Anti de Sitter Space'', JHEP  0006 (2000) 008 [hep-th/0003075].
\item  M. T. Grisaru, R. C. Myers and O. Tafjord, ``SUSY and Goliath'', JHEP  0008 (2000) 040 [hep-th/0008015]; A. Hashimoto, S. Hirano and N. Itzakhi, ``Large Branes in AdS and their Field Theory Dual," 0008 (2000) 051 (2000)  [hep-th/0008016].
\item  V. Balasubramanian, M. Berkooz, A. Naqvi and M. Strassler", Giant Gravitons in Conformal Field Theory", JHEP 0204 (2002) 034 [hep-th/0107119];
S. Corley, A. Jevicki and S. Ramgoolam, ``Exact Correlators of Giant Gravitons from Dual N = 4 SYM Theory,'' Adv. Theor. Math. Phys. 5 (2002) 809 
2002  [hep-th/0111222]; S. Corley and S. Ramgoolam, ``Finite Factorization equations and sum rules for BPS Correlators in N=4 SYM Theory,'' Nucl. Phys. B641 (2002) 131  [hep-th/0205221];  D. Berenstein, ``Shape and Holography: Studies of dual operators to giant gravitons," Nucl. Phys. B675 (2003) 179 [hep-th/0306090];  R. de M. Koch and R. Gwyn, ``Giant Graviton Correlators from Dual $SU(N)$ super Yang-Mills Theory," JHEP 0411 (2004) 081 [hep-th/0410236]. 
\item O. Lunin, S. D. Mathur, and A. Saxena, ``What is the gravity dual of a chiral primary?,''  Nucl.Phys. B655 (2003) 185 [hep-th/0211292].
\item  O.~Lunin and J.~Maldacena, ``Deforming field theories with U(1) x U(1) global symmetry and their gravity duals,'' JHEP  0505  (2005)  033  [hep-th/0502086]. 
\item R.~G.~Leigh and M.~J.~Strassler, ``Exactly marginal operators and duality in four-dimensional N=1 supersymmetric gauge theory,'' Nucl.\ Phys.\ B447 (1995) 95 [hep-th/9503121].
\item  Wung-Hong Huang, ``Semiclassical Rotating Strings in  Electric and Magnetic Fields Deformed $AdS_5 \times S^5$ Spacetime'', Phys. Rev. D73 (2006) 026007 [hep-th/0512117].  
\item  Wung-Hong Huang, ``Electric/Magnetic Field Deformed Giant Gravitons in Melvin Geometry'', Phys.Lett. B635 (2006) 141 [hep-th/0602019 ].  
\item B. Janssen, Y. Lozano, and D. Rodriguez-Gomez,``Giant Gravitons in $AdS_3 \times S^3 \times T^4$ as Fuzzy Cylinders,'' Nucl.Phys. B711 (2005) 392 [hep-th/0406148].
\item M.A. Melvin, ``Pure magnetic and electric geons,'' Phys. Lett. 8 (1964) 65; F. Dowker, J. P. Gauntlett, D. A. Kastor and Jennie Traschen, ``Pair Creation of Dilaton Black Holes,'' Phys.Rev. D49 (1994) 2909-2917  [hep-th/9309075]; F.~Dowker, J.~P.~Gauntlett, D.~A.~Kastor and J.~Traschen, ``The decay of magnetic fields in Kaluza-Klein theory,'' Phys.\ Rev.\ D52 (1995) 6929 [hep-th/9507143]; M.~S.~Costa and M.~Gutperle, ``The Kaluza-Klein Melvin solution in M-theory,'' JHEP 0103 (2001) 027 [hep-th/0012072].
\item A. A. Tseytin, ``Harmonic superpositions of M-brnaes,'' Nucl. Phys. B475 (1996) 149  [hep-th/9604035];  J. P. Gauntlett, D. A. Kastor, and J. Traschen, ``Overlapping Branes in M-Theory,'' Nucl. Phys. B478 (1996) 149  [hep-th/9604179].
\item E. Bergshoeff, C.M. Hull, T. Ortin,, ``Duality in the Type-II Superstring Effective Action,'' Nucl.Phys. B451 (1995) 547 [hep-th/9504081];  S. F. Hassan, ``T-Duality, Space-time Spinors and R-R Fields in Curved Backgrounds,'' Nucl.Phys. B568 (2000) 145 [hep-th/9907152].
\item G.~W.~Gibbons and D.~L.~Wiltshire, ``Space-time as a membrane in higher dimensions,'' Nucl.\ Phys.\ B287 (1987) 717 [hep-th/0109093]; G.~W.~Gibbons and K.~Maeda, ``Black holes and membranes in higher dimensional theories with dilaton fields,'' Nucl.\ Phys.\ B298 (1988) 741; L. Cornalba and M.S. Costa, ``A New Cosmological Scenario in String Theory,'' Phys.Rev. D66 (2002) 066001 [hep-th/0203031]; T. Friedmann and H. Verlinde, ``Schwinger pair creation of Kaluza-Klein particles: Pair creation without tunneling,'' Phys.Rev. D71 (2005) 064018 [hep-th/0212163].
\item  S. R. Das, S. P.. Trivedi, S. Vaidya,``Magnetic Moments of Branes and Giant Gravitons,''  JHEP 0010 (2000) 037 [hep-th/0008203].
\item J.~M.~Maldacena and L. Maoz, ``De-singularization by rotation,'' [hep-th/0012025]; V. Balasubramanian, J. de Boer, E. Keski-Vakkuri, S. F. Ross, ``  Supersymmetric Conical Defects: Towards a string theoretic description of black hole formation,'' Phys.Rev. D64 (2001) 064011 [hep-th/0011217].
\item M. Cvetic,and D. Youm, ``Rotating Intersecting M-Branes,''  Nucl.Phys. B499 (1997) 253 [hep-th/9612229].  
\end{enumerate}
\end{document}